\documentclass[12pt]{iopart}
\usepackage{amsbsy}
\usepackage{graphicx, epsfig, amssymb} 
\usepackage{bm} 

\usepackage[linktocpage]{hyperref}
\usepackage[caption=false]{subfig}
\usepackage[usenames]{color}
\usepackage{url}
\usepackage{color}

\newcommand{\be}{\begin{equation}}
\newcommand{\ee}{\end{equation}}
\newcommand{\beq}{\begin{eqnarray}}
\newcommand{\eeq}{\end{eqnarray}}

\begin{document}

\title[Raising tides on black holes]{Tidal acceleration of black holes and superradiance}


\author{Vitor Cardoso\footnote{vitor.cardoso@ist.utl.pt}$^{1,2}$, Paolo Pani\footnote{paolo.pani@ist.utl.pt}$^{1}$}

\address{$^{1}$ Centro Multidisciplinar de Astrof\'\i sica --- CENTRA, Departamento de F\'\i sica, Instituto Superior T\'ecnico --- IST,
Universidade T\'ecnica de Lisboa - UTL, Av. Rovisco Pais 1, 1049-001 Lisboa, Portugal}
\address{$^{2}$ Department of Physics and Astronomy, The University of Mississippi,
  University, MS 38677-1848, USA}

\begin{abstract}
Tidal effects have long ago locked the Moon in synchronous rotation with the Earth and progressively increase the Earth-Moon distance. This \emph{``tidal acceleration''} hinges on dissipation. Binaries containing black holes may also be tidally accelerated, dissipation being caused by the event horizon -- a flexible, viscous one-way membrane. 
In fact, this process is known for many years under a different guise: \emph{superradiance}.  Here we provide compelling evidence for a strong connection between tidal acceleration and superradiant scattering around spinning black holes. In General Relativity, tidal acceleration is obscured by gravitational-wave emission. However, when coupling to light scalar degrees of freedom is allowed, an induced dipole moment produces a \emph{``polarization acceleration''}, which might be orders of magnitude stronger than tidal quadrupolar effects. Consequences for optical and gravitational-wave observations are intriguing and it is not impossible that imprints of such mechanism have already been observed.
\end{abstract}

\maketitle

\markboth{Tidal acceleration of black holes and superradiance}{Tidal acceleration of black holes and superradiance}

\section{Introduction}
%
\begin{figure}
\begin{center}
\begin{tabular}{cc}
\epsfig{file=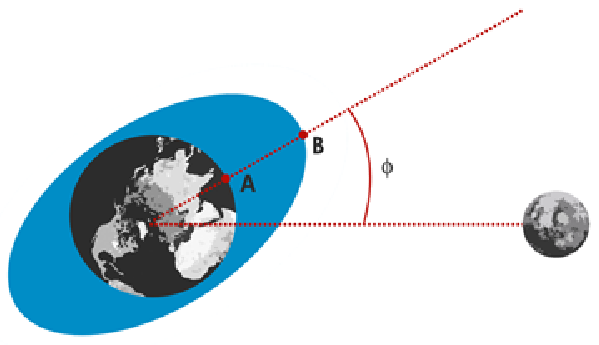,width=7cm,angle=0,clip=true}&
\epsfig{file=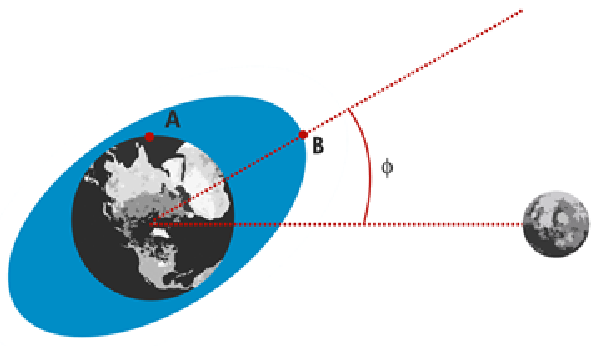,width=7cm,angle=0,clip=true}
\end{tabular}
\caption{Tides on the Earth caused by our moon (as seen by a frame anchored on the moon). The tidal forces create a bulge in the Earth, which leads the moon in its orbit by a constant angle $\phi$. The Earth rotates faster than the moon in its orbit, thus a point $A$ on the surface of the Earth will differentially rotate with respect to the oceans, causing dissipation of energy and decrease of the Earth's rotation period.
\label{fig:tides}}
\end{center}
\end{figure}
The gravitational pull of the Moon upon the Earth produces tides and, because Earth rotates, there are two tides a day. 
Tidal effects are also responsible for the constant drift of Moon's orbit -- known as \emph{tidal acceleration} -- and for its synchronous rotation with Earth, called \emph{tidal locking}. These fascinating phenomena have been influential in astrophysics, space sciences and also in the arts, by inspiring for example Italo Calvino's Cosmicomic, ``The Distance of the Moon'' and  Pink Floyd's masterpiece  ``The Dark Side of the Moon''.

In 1880, George Darwin explained the basic mechanism of how tides work~\cite{Darwin:1880}. 
Excellent expositions can be found in the reviews by Hut~\cite{Hut:1981} and Verbunt~\cite{Verbunt}. Tides are caused by differential forces on the oceans, and they raise tidal bulges on them, as depicted in Figure~\ref{fig:tides}. Because Earth rotates with angular velocity $\Omega_{\rm Earth}$, these bulges are not exactly aligned with the Earth-Moon direction. In fact, because Earth rotates faster than the Moon's orbital motion ($\Omega_{\rm Earth}>\Omega$), the bulges {\it lead} the Earth-Moon direction by a constant angle. This angle would be zero if friction were absent, and the magnitude of the angle depends on the amount of friction. Friction between the ocean and the Earth's crust slows down Earth's rotation by roughly $\dot{\Omega}_{\rm Earth}\sim -5.6\times 10^{-22}/s^2$, about $0.002s$ per century. Conservation of angular momentum of the entire system lifts the Moon into a higher orbit with a longer period and larger semi-major axis (lunar ranging estimates put this at around $\dot{a}=3.82 {\rm cm}/{\rm yr}$~\cite{Dickey:1994zz}). 

A complete understanding of tides cannot be obtained without a detailed knowledge of the physics behind the dissipation mechanism which, for the Earth-Moon system, is not well understood and can -- at best -- be constrained through observational data. Fortunately, when the tidally distorted object is a black hole (BH), the physics and mathematics behind dissipation are under control. 
Indeed, in several cases of astrophysical relevance the moon can be treated as a small perturbation of an isolated, stationary BHs. The no-hair theorems of General Relativity (see Ref.~\cite{Bekenstein:1996pn} for a review) guarantee that the background is described by the Kerr metric and it is solely characterized by two parameters: its mass and angular momentum.
The first important step towards understanding BH tides
was given by Hartle many years ago \cite{Hartle:1973zz,Hartle:1974gy}. More recently, Poisson and co-workers \cite{Poisson:2009di,Binnington:2009bb} established an elegant correspondence between BH tides and those raised on viscous fluids. In this paper, we shall provide compelling evidence for a broader correspondence between tidal acceleration and superradiant scattering~\cite{zeldovich,Teukolsky:1973ha,Bekenstein:1973mi} around a spinning BH. Our results extend the membrane paradigm~\cite{membranep} by providing a novel and simple understanding of some properties of the event horizon and, as we shall discuss, they might be useful to interpret some elusive features of BH dynamics.

\section{Tides in Newtonian gravity}
Let us consider a generic power-law interaction between a central body of gravitational mass $M$ and radius $R$ and its moon with mass $m_p$ at a distance $r_0$. The magnitude is
\be
F=\frac{GMm_p}{r_0^{n}}\,, \label{NewtonGeneral}
\ee
%
%
%
and Newton's law is recovered for $n=2$.
The tidal acceleration in $M$ is given by
\be
a_{\rm tidal}=\frac{nGm_p}{R^n}\left(\frac{R}{r_0}\right)^{n+1}=n g_M\left(\frac{R}{r_0}\right)^{n+1}\frac{m_p}{M}\,,
\ee
where $g_M$ is the surface gravity on $M$.
This acceleration causes tidal bulges of height $h$ and mass $\mu$ to be raised on $M$. These can be estimated by equating the specific energy of the tidal field, $E_{\rm tidal}\sim a_{\rm tidal} R$, with the specific gravitational energy, $E_G\sim g_M h$, needed to lift a unit mass from the surface of $M$ to a distance $h$. We get
\be
\frac{h}{R}=n\left(\frac{R}{r_0}\right)^{n+1}\frac{m_p}{M}\,,
\ee
which corresponds to a bulge mass of approximately
\be
\mu=\frac{\kappa}{4}n m_p \left(\frac{R}{r_0}\right)^{n+1}\,.
\ee
where $\kappa$ is a constant of order unity, which encodes the details of Earth's internal structure.
Without dissipation, the position angle $\phi$ in Figure~\ref{fig:tides} is $\phi=0$, while the tidal bulge is aligned with moon's motion. Dissipation contributes a constant, small, time lag $\tau$ such that the lag angle is
\be
\phi=(\Omega_{\rm Earth}-\Omega)\tau\,.
\ee
With these preliminaries, a trivial extension of Hut's~\cite{Hut:1981} result yields a tangential tidal force on $M$ (we assume a circular orbit for the moon) 
\be
F_{\theta}\sim \frac{n(n+1)G\kappa}{2}m_p^2\frac{R^{n+3}}{r_0^{2n+3}}(\Omega_{\rm Earth}-\Omega)\tau\,.
\ee
This perturbation exerts a torque $N=r_0F_{\theta}$. The change in orbital energy over one orbit is
$\int_0^{2\pi}r_0F_{\theta}\Omega/2\pi d\theta=\Omega r_0 F_{\theta}$. Thus, we get
\be
\dot{E}_{\rm orbital}=\frac{n(n+1)G\kappa m_p^2}{2}\frac{R^{n+3}}{r_0^{2n+2}}\Omega(\Omega_{\rm Earth}-\Omega)\tau\,. \label{tidesGeneral}
\ee
For gravitational forces obeying Gauss's law ($n=2$) this yields
\be
\dot{E}_{\rm orbital}=3G\kappa m_p^2\frac{R^{5}}{r_0^{6}}\Omega(\Omega_{\rm Earth}-\Omega)\tau\,.\label{tidesNewt}
\ee
%
\section{No tidal acceleration of black holes in General Relativity}
Tidal interactions between BHs and moons were investigated many years ago by Hartle~\cite{Hartle:1973zz,Hartle:1974gy} and more recently by Poisson and co-workers~\cite{Poisson:2009di,Binnington:2009bb}. A moon of mass $m_p$ orbiting a BH of mass $M$ and angular velocity $\Omega_{\rm BH}$ at a distance $r_0$ with orbital frequency $\Omega$, dissipates energy (through tidal heating) at the event horizon at a rate of
\be
\dot{E}_{\rm H}\sim \frac{G^7}{c^{13}}\frac{M^6m_p^2}{r_0^6}\Omega (\Omega-\Omega_{\rm BH})\,.\label{horizontides}
\ee
This result can be derived within BH perturbation theory in the extreme mass ratio limit~\cite{Press:1972zz}, when the energy fluxes generated by a point-particle orbiting a spinning BH at orbital frequency $\Omega$ are computed within the Teukolsky formalism~\cite{Teukolsky:1973ha}. 

Our crucial point is to recognize that Eq.~(\ref{horizontides}) follows from (\ref{tidesNewt}) by substituting $\Omega_{\rm Earth}\to\Omega_{\rm BH}$, setting $\kappa\sim1/3\approx{\cal O}(1)$ and with the simple argument that the only relevant timescale in the BH case is a light-crossing time, $\tau\sim R/c$, where $R=GM/c^2$. 
It is remarkable that our Newtonian computation is in agreement with a much more involved perturbative analysis of Einstein's equations at fully relativistic level~\cite{Poisson:1994yf}.

If this was the end of the story, BHs could be spun-down by the moon, since for 
\begin{equation}
 \Omega<\Omega_{\rm BH} \,,\label{condtidal}
\end{equation}
energy is flowing {\it out} of the BH. The BH spins down and angular momentum conservation imposes that the moon spirals outwards over secular timescales. However, this is not the end of the story, because tidal heating is small and BHs are General Relativistic objects made by pure spacetime fabric. Any tidal distortion also carries energy (under gravitational waves) to infinity, at a quadrupolar rate $|\dot{E}_{\infty}|\sim 32 G^4 M^3 m_p^2 /(5 c^5 r_0^5)$.
%
%
Thus, a net tidal acceleration, which we will define here as spinning down the central BH and pushing {\it outwards} its moon, is only possible if the rate at which energy is dissipated to infinity is smaller than the rate at which energy flows out of the BH, $|\dot{E}_{\rm H}|/\dot{E_{\infty}}>1$. This condition is paramount to
\be
\frac{|\dot{E}_{\rm H}|}{\dot{E}_{\infty}}=\left(\frac{GM}{c^2r_0}\right)^3 \frac{r_0^2\Omega}{c^2}\left({\Omega-\Omega_{\rm BH}}\right)\sim \left({v}/{c}\right)^5\gtrsim 1\,,\label{ratio_tidal}
\ee
where we assumed $\Omega\ll\Omega_{\rm BH}$ and $v=(M\Omega)^{1/3}\sim\sqrt{M/r_0}$ is the moon orbital velocity. While our analysis provides a simple Newtonian understanding of this relation above, nonetheless the inequality above can never be satisfied in General Relativity. 
BH tidal acceleration in Einstein's theory is impossible because tidal effects are completely washed out by gravitational-wave emission. As a consequence, moons always spiral inwards BHs and are eventually swallowed.
\section{Another take on tidal acceleration: superradiance}
The agreement between Eq.~(\ref{ratio_tidal}) and more detailed computations involving BH perturbations \cite{Poisson:1994yf} suggests a connection between two different perspectives: tidal absorption and heating at the horizon on one side, wave absorption and BH perturbations on the other. In particular, the fact that the energy flux at the horizon is \emph{negative} when $\Omega<\Omega_H$  is analogous to the well-known \emph{superradiance} condition~\cite{zeldovich,Teukolsky:1973ha,Bekenstein:1973mi}. A wave scattered off a spinning BH is superradiantly amplified if $\omega<m\Omega_H$, where $\omega$ is the frequency of the wave and $m$ is the azimuthal number. Particles in circular orbits emit waves at the frequency $\omega=m\Omega$. Therefore, in the inspiral of a point-particle around a spinning BH, superradiance occurs precisely when the condition~(\ref{condtidal}) for tidal acceleration is met. In order to elaborate on this analogy further, in this and in the next sections we shall show that an equivalent, and complementary, approach to understand BH tidal acceleration can be provided solely from a wave-like perspective and that superradiance is the wave analog of Newtonian tidal heating.

Looking at the orbiting moon as a small time-dependent disturbance in a stationary rotating background, the full spacetime can be described in terms of wave-like quantities, such as Newman-Penrose scalars. Massless fields propagating near a rotating BH can be expressed in terms of a single master variable $\Psi$ which obeys a Schrodinger-type equation of the form~\cite{Teukolsky:1973ha,arXiv:0905.2975}
\be \frac{d^2 \Psi}{dz_*^2}-V_{\rm eff} \Psi=0\,,\label{wave} 
\ee 
where we have defined the tortoise coordinate $z_*$ which covers the entire real axis.
In a scattering experiment of a wave with frequency $\omega$ and azimuthal and time dependence $e^{-i\omega t+im\varphi}$, equation (\ref{wave}) has the following asymptotic behavior 
%
\be
\Psi _1 \sim\left\{
\begin{array}{ll}
{\cal T}(r-r_H)^{-i\chi\left(\omega-m\Omega_{\rm BH}\right)}+{\cal O}(r-r_H)^{i\chi\left(\omega-m\Omega_{\rm BH}\right)} & {\rm as}\ r\rightarrow r_H \,, \\
{\cal R}{\rm e}^{i\omega r}+ {\rm e}^{-i \omega r}& {\rm as}\ r\rightarrow \infty\,.
\end{array}
\right. \label{bound2}
\ee
where $r_H$ is the horizon radius in some chosen coordinates, $\Omega_{BH}$ is the angular velocity at $r=r_H$ of locally nonrotating observers, $\chi$ is a parameter that depends on the specific rotating background and, for a Kerr BH\footnote{Note that our argument is only based on the existence of a spinning, asymptotically-flat BH and on the asymptotic behaviors of the perturbations. In particular, the argument is valid also for BH metrics other than Kerr.}, it reads $\chi =2r_H(r_H^2+a^2)/(r_H^2-a^2)$, where $a$ is the angular momentum per unit mass,
$a/M\equiv c J/(G M^2)$.
These boundary conditions correspond to an incident wave of unit amplitude from spatial infinity giving rise to a reflected wave of amplitude ${\cal R}$ and a transmitted wave of amplitude ${\cal T}$ at the horizon. 
The ${\cal O}$ term describes a putative out-going flux across the surface at $r=r_H$. The presence of a horizon and a well-posed Cauchy problem would imply ${\cal O}\equiv0$. Here we shall generically keep this term, in order to allow for a nonvanishing out-going flux in absence of an event horizon.

Let us assume that there is no dissipation mechanism other than -- possibly -- a dissipative membrane at $r=r_H$. Then, the potential is real and the complex conjugate $\Psi_2=\bar{\Psi}_1$ will satisfy the complex-conjugate boundary conditions. 
The solutions $\Psi_1$ and $\Psi_2$ are linearly independent, and standard theory of ODEs tells us that their Wronskian is a constant
(independent of $r$). If we evaluate the Wronskian near the horizon, we get $W= -2i\left(\omega-m\Omega_{\rm BH}\right)\left(|{\cal T}|^2-|{\cal O}|^2\right)$. On the other hand, at infinity we have $W=2i \omega(|{\cal R}|^2-1)$. Equating the two we get 
\be 
|{\cal R}|^2=1-\frac{\omega-m\Omega_{\rm BH}}{\omega}\left(|{\cal T}|^2-|{\cal O}|^2\right)\,,\label{wronskian}
\ee 
independently from the details of the potential in wave equation.
There are some important remarks to be made about the relation above.
Let us consider the equation above in the case of a one-way membrane boundary conditions at the horizon, i.e. ${\cal O}=0$.
In general $|{\cal R}|^2<1$, as is to be expected for scattering off perfect absorbers. However, for 
$\omega-m\Omega_{\rm BH}<0$, we have a \emph{superradiant} regime $|{\cal R}|^2>1$~\cite{Teukolsky:1974yv}. The excess energy comes from the hole's rotational energy, which therefore decreases. 

Furthermore, notice how {\it dissipation} is a crucial ingredient for superradiance: without in-going boundary conditions at the horizon, no superradiant scattering can occur, as discussed in Refs.~\cite{zeldovich,Bekenstein:1973mi,Richartz:2009mi}. In absence of a horizon (for example in the case of rotating perfect-fluid stars), regularity boundary conditions must be imposed at the center of the object. By applying the same argument as above, the Wronskian at the center vanishes, which implies $|{\cal R}|^2=1$, i.e. no superradiance. If the rotating object does not possess a horizon, superradiance can only come from some other dissipation mechanism, like friction due to oceans or to the atmosphere, which anyway require a precise knowledge of the microphysics governing the interior of the object.
Equivalently, we can argue that $|{\cal O}|^2$ and $|{\cal T}|^2$ are respectively proportional to the outgoing and transmitted energy flux across the surface at $r_H$. In absence of dissipation, energy conservation implies that the out-going flux will equal the transmitted one, i.e. $|{\cal O}|^2=|{\cal T}|^2$ and Eq.~(\ref{wronskian}) would again prevent superradiance, $|{\cal R}|^2=1$.



%
\begin{figure}
\begin{center}
\begin{tabular}{cc}
\epsfig{file=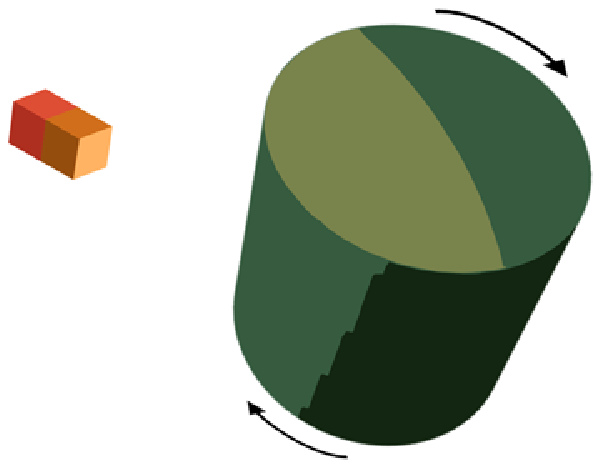,width=4.3cm,angle=0,clip=true}&
\epsfig{file=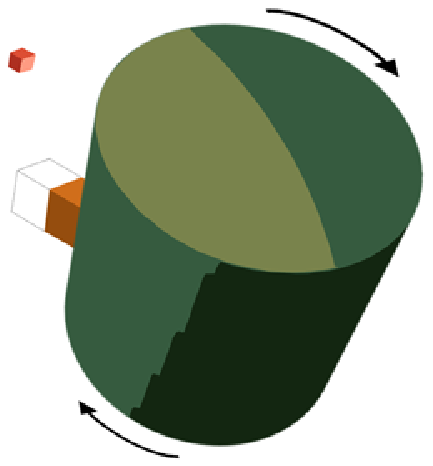,width=4.3cm,angle=0,clip=true}
\end{tabular}
\caption{Pictorial description of a classical analog of the Penrose process. A body falls nearly from rest into a rotating cylinder, whose surface is sprayed with glue. At the surface the body is forced to co-rotate with the cylinder (analog therefore of the BH ergosphere, the surface beyond which no observer can remain stationary with respect to infinity). The negative energy states of the ergoregion are played by the potential energy associated with the sticky surface. If now half the object (in reddish) is detached from the first half (yellowish), it will reach infinity with more (kinetic) energy than it had initially, extracting rotational energy out of the system. 
\label{fig:ergoregion}}
\end{center}
\end{figure}

We have therefore established that, for spinning and stationary spacetimes in absence of some dissipation mechanism other than (possibly) a one-way membrane, tidal acceleration of small moons and superradiant scattering of probe fields come hand-in-hand: they can occur \emph{if and only if} the rotating background metric possesses an event horizon. In particular, BH tidal acceleration cannot occur without superradiance and viceversa. A particle orbiting a spinning BH with orbital frequency $\Omega$ raises tides on the horizon if $\Omega<\Omega_H$. In the wave counterpart, the gravitational perturbations sourced by the particle generate a negative energy flux across the horizon when the superradiant condition is met, $\Omega<\Omega_H$. More precisely, the tidal heating computed at Newtonian level [cf. Eq.~(\ref{horizontides})] agrees with the energy flux extracted from the horizon via superradiant scattering as obtained by solving Einstein's equations perturbatively~\cite{Teukolsky:1974yv,Poisson:1994yf}.
In fact, these effects are two sides of the same coin. The underlying reason is due to purely gravitational nature of the dissipation mechanism for isolated BHs, which is solely provided by the one-way membrane behavior of the horizon. The very same mechanism is responsible for the amplification of scattered fields around a spinning BH.

%
\subsection*{Tidal acceleration, superradiance and Penrose process}
In order to elaborate on this connection further, it is instructive to consider the case of a spinning object with no horizon. In this case, there exists another classical process  devised by Penrose in order to extract angular momentum from the object~\cite{Penrose:1969pc}. Penrose's process (see Figure \ref{fig:ergoregion} for a flat-space, classical analog) is tightly connected with the existence of negative energy regions -- so called \emph{ergoregions} -- around spinning relativistic metrics. In brief, an incident particle disintegrates into two fragments within the ergosphere: one fragment can have negative energy and be confined inside the ergoregion, whereas the other fragment can reach infinity with a larger energy than the original incident particle. By energy conservation, this amplification implies energy extraction from the central spinning object~\cite{Penrose:1969pc}. This process does \emph{not} require an event horizon and one would be tempted to conclude that the same applies to superradiance. However, in the case of superradiance the existence of an ergoregion alone (for example as around highly spinning stars) is not sufficient~\cite{Richartz:2009mi}: as we have shown above dissipation, as provided by an event horizon, is another crucial ingredient. The role of the ergoregion in this correspondence is to account for exchange of energy between the BH and the moon, even in a fixed background approximation. 
In fact, it is impossible to disentangle the role of the ergoregion from that of the event horizon because, as we prove in the Appendix, an ergoregion necessarily exists in the exterior of any stationary and axisymmetric BH. As a by-product, superradiance (and therefore tidal heating) is a sufficient condition for the Penrose process to occur, but the converse is not necessarily true.

\section{Tidal or polarization effects in the presence of long-range fields}
%
\begin{figure}
\begin{center}
\begin{tabular}{cc}
\epsfig{file=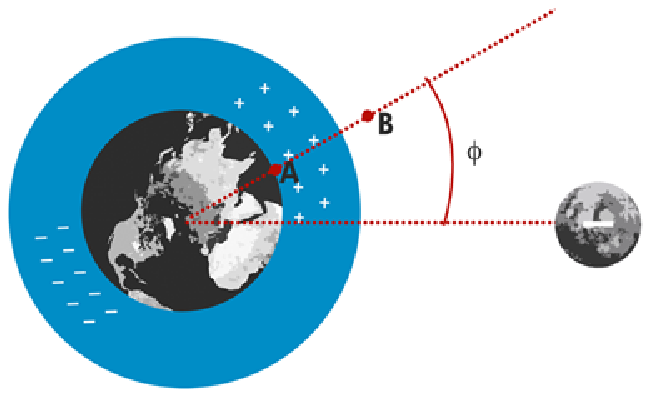,width=7cm,angle=0,clip=true}&
\epsfig{file=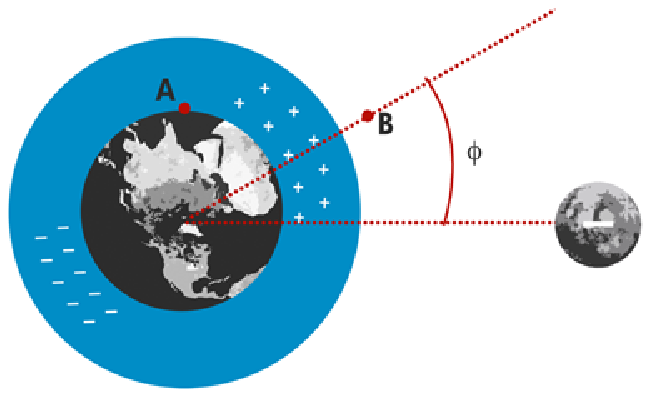,width=7cm,angle=0,clip=true}
\end{tabular}
\caption{``Tides'' or polarization of the Earth caused by an electrically charged moon (as seen by a frame anchored on the moon). The moon's electric field creates bulges of surface charge on the Earth, imparting to the (electrically neutral) Earth
a dipole moment. The charged bulges lead the moon in its orbit by a constant angle $\phi$, on account of the Earth rotating faster that the moon's orbital motion. 
\label{fig:polarization}}
\end{center}
\end{figure}
In this and in the next section we shall extend the gravitational analogy between BH tidal acceleration and BH superradiance to the case of electromagnetic and scalar interactions. 
In the present context, the motivation for studying such long-range fields is twofold. First, from a purely theoretical standpoint, by allowing for more general interactions we shall provide further support to the connection we are trying to establish. Secondly, from a more phenomenological viewpoint, we shall show that other types of interactions can give rise to much stronger tides and, in some scenarios, tidal effects can even dominate over gravitational-wave emission at infinity, providing a distinctive signature of BH tidal heating.

Einstein's theory is one of mankind's greatest achievements, yet it is thought to be an incomplete theory, incompatible with the Standard Model of particle physics. Several attempts at a unified theory (such as string theory) predict the existence of a plethora of extra, light scalar fields adding to Einstein's original description of gravity~\cite{Arvanitaki:2009fg,Arvanitaki:2010sy}. Such fields seem to be a quite generic feature of modified theories of gravity and they are being looked for in a variety of experiments. They naturally arise in several modified gravities, such as scalar-tensor theories of which Brans-Dicke's is an example. Furthermore, they are a very convenient proxy for other more complex interactions and often arise as effective degrees of freedom, for instance in $f(R)$ theories~\cite{DeFelice:2010aj}. Therefore, beside considering standard electromagnetic fields, in the following we shall focus on scalar-type interactions coupled to gravity. By providing an equivalent description of this effect in terms of BH superradiance, we shall describe how these scalar fields affect the dissipation mechanism and raise tides on the BH horizon.
\subsection{Newtonian ``polarization'' acceleration or tides for charged interactions}
Gravity is a very special interaction because there are no negatively
charged particles and, by the equivalence principle, all objects fall in the same way. Although (seemingly) not previously studied, charged fields also raise tides on BHs. If one of the interacting bodies is charged while the other is neutral, polarization of the neutral body will occur, as depicted in Figure~\ref{fig:polarization}. As we prove below, polarization leads to a \emph{dipolar} tidal acceleration which is stronger than its gravitational counterpart. We call this leading-order dipole effect a ``polarization acceleration'' to distinguish it from the standard quadrupolar gravitational tides.

Let us then consider the interaction of a \emph{charged} moon with (either electric or scalar) charge $q_p$ and mass $m_p$ with a neutral central object of mass $M$ and size $R$, a distance $r_0$ apart. 
It is easy to see that a splitting of induced charges in the central object will make tidal effects subdominant with respect to ``polarization'' effects. If the object has a dielectric constant $\epsilon=\epsilon_r \epsilon_0$, the moon's external field induces a polarization surface charge density and dipole moment in the central object of respectively~\cite{Jackson}
\be
\sigma_{\rm pol}=3\epsilon_0\left(\frac{\epsilon_r-1}{2\epsilon_r+1}\right)E_0 \cos\theta\,,\qquad
p=4\pi \epsilon_0\left(\frac{\epsilon_r-1}{2\epsilon_r+1}\right)R^3E_0 \,,\label{surfacepol}
\ee
where $E_0=q_p/(4\pi\epsilon_0r_0^2)$ and $\theta$ is a polar angle around the massive object. The induced electric field at the location of $q_p$ is then
\be
\vec{E}=\frac{3\vec{n}(\vec{p}\cdot\vec{n})-\vec{p}}{4\pi\epsilon_0 r_0^3}
\ee
with $\vec{n}$ a unit vector directed from $M$ to $q_p$. Without dissipation, the position angle is $\phi=0$, while the tidal bulge is aligned with the moon's motion. Again, dissipation contributes a constant, small time lag $\tau$, such that the lag angle is
$\phi=(\Omega_{\rm Earth}-\Omega)\tau$. The tangential force is now
\be
F_{\theta}\sim \frac{q_p p}{4\pi\epsilon_0 r_0^3}(\Omega_{\rm Earth}-\Omega)\tau\,,
\ee
while the change in orbital energy over one orbit is
\be
\dot{E}_{\rm orbital}=\Omega r_0 F_{\theta}=\left(\frac{\epsilon_r-1}{2\epsilon_r+1}\right)\frac{q_p^2R^3 \tau}{r_0^4}\Omega(\Omega_{\rm Earth}-\Omega)\,.\label{Edotpol}
\ee
%
\subsection{``Polarization'' acceleration of black holes}
Polarization acceleration when the central object is a BH can be explained in a manner similar to what we previously discussed for tidal effects.
It has long been known that the event horizon acts as a one-way membrane with a certain characteristic resistivity, viscosity
and electric permittivity. In this ``membrane paradigm''~\cite{membranep} BHs, when acted upon by an external field due to the presence of a point charge, acquire a surface density
$\sigma=q_p \cos\theta/(4\pi r_0^2)$ for $r_0 \gg GM/c^2$. Interestingly, by comparing this result with Eqs.~(\ref{surfacepol}) we obtain that BHs have precisely $\epsilon_r=4$, a relative permittivity similar to that of paper. Thus, from Eq.~(\ref{Edotpol}) we get
\be
|\dot{E}_{\rm H}|=\frac{G^4q_p^2}{3c^9}\frac{M^4}{r_0^4}\Omega(\Omega-\Omega_{\rm BH})\,. \label{EdotHpol}
\ee
where we substituted $\Omega_{\rm Earth}\to\Omega_{\rm BH}$ and we assumed again $\tau\sim GM/c^3$.
On the other hand, (electrically) charged moons radiate according to Larmor's formula, $\dot{E}_{\infty}=q_p^2 \Omega^4r_0^2/(6\pi\epsilon_0 c^3)$.
Our Newtonian analysis, together with the membrane paradigm, predicts the following flux ratio:
\be
\frac{|\dot{E}_{\rm H}|}{\dot{E}_{\infty}}=\frac{2\pi \epsilon_0 G^4}{c^6}\frac{M^4}{r_0^6\Omega^3}(\Omega-\Omega_{\rm BH})\sim (v/c)^3\,,\label{ratiocharge}
\ee
where in the last equation we assumed $\Omega\ll\Omega_{\rm BH}$. In fact, this expression holds for any Gauss-law interaction, in particular for both electromagnetic and massless scalar interactions. Noteworthy, Eq.~(\ref{ratiocharge}) can be shown to be in agreement with relativistic predictions coming from an analysis of \emph{Maxwell} equations~\cite{Goldberger:2005cd,Porto:2008}, and it also agrees with numerical computations of the BH \emph{scalar} dipolar emission in scalar-tensor theories~\cite{Cardoso:2011xi,Yunes:2011aa}. To be more specific, the flux ratio~(\ref{ratiocharge}) is, to leading order, consistent with the flux generated by a scalar point charge orbiting a spinning BH with orbital frequency $\Omega$. The latter computation is based on a Teukolsky decomposition~\cite{Teukolsky:1973ha} of the gravitational and scalar perturbations in the Fourier domain. It is remarkable that the same result can be obtained and understood just in terms of Newtonian polarization acceleration. Similarly to the gravitational case described above, the energy flux across the horizon is negative when the superradiance condition is satisfied, $\Omega-\Omega_H<0$, which is consistent with Eq.~(\ref{EdotHpol}).
To our knowledge, this is the first time that a direct association is made between energy fluxes across the BH horizon and polarization tides raised at the horizon by an orbiting charged moon.
\section{Polarization acceleration in the presence of massive scalar fields}
\begin{figure}
\begin{center}
\begin{tabular}{cc}
\epsfig{file=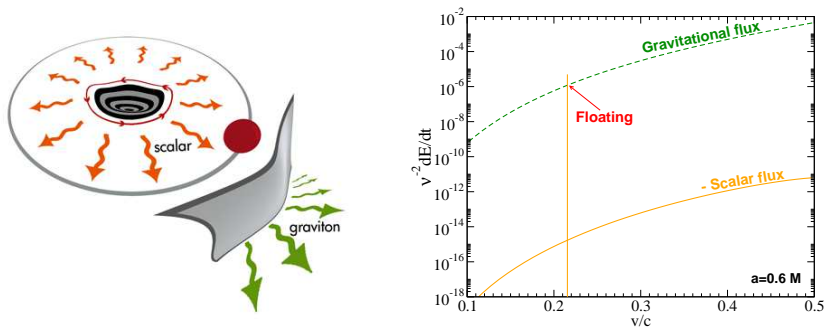,width=5.5cm,angle=0,clip=true}&
\epsfig{file=Plots/floating.eps,width=5.5cm,angle=0,clip=true}
\end{tabular}
\caption{Pictorial description of floating orbits. An orbiting body excites superradiant scalar modes close to the BH horizon (straight orange line on the right). Since the scalar field is massive, the flux at infinity consists solely of gravitational radiation (dashed green line on the right). Superradiant resonances excite the scalar flux at the horizon to (absolute) values which may be larger than the gravitational flux. The orbiting particle is driven to ``floating orbits'' for which the orbital velocity is such that the total flux is vanishing (see Refs.~\cite{Cardoso:2011xi,Yunes:2011aa} for details).
\label{fig:superradiance2}}
\end{center}
\end{figure}

We have so far discussed two important issues: (i) BH tidal acceleration can be understood in terms of superradiance for gravitational, electromagnetic and massless scalar interactions and (ii) dipole ``polarization'' effects are dominant over tidal quadrupoles, e.g. compare the small-$v$ behavior of Eq.~(\ref{ratiocharge}) with that of Eq.~(\ref{ratio_tidal}). However, even in the case of polarization acceleration, from Eq.~(\ref{ratiocharge}) we get ${|\dot{E}_{\rm H}|}/{\dot{E}_{\infty}}\to0$ when $v\ll c$, i.e. tidal effects are still too weak to compensate for wave emission at infinity. With these results at hand, we can now discuss instances in which polarization effects are enormously amplified.
This is the case when superradiant modes of a rotating BH are \emph{excited} by the orbital motion of its moon. 
In General Relativity, the ringing modes of BHs correspond roughly to the light-ring frequencies~\cite{arXiv:0905.2975} and they lay inside the innermost stable circular orbit (ISCO), thus preventing the possibility of resonances due to moon's orbital motion. However, when coupling to light scalar degrees of freedom is allowed, a different class of modes can be excited, as depicted in Figure~\ref{fig:superradiance2}. 
For example, the characteristic modes of scalar massive perturbations around a Kerr BH scale with the mass of the scalar field, $\mu_s$. In the small mass limit the real and imaginary part of the mode read~\cite{Detweiler:1980uk},
\begin{equation}
 \omega_R\sim \mu_s\,,\qquad \omega_I M\sim \mu_s^9 M^9\,. \label{wRwI}
\end{equation}
Just as a damped and forced oscillator, these resonant frequencies can be excited by point-like particles moving along stable circular orbits, when the orbital frequency equals the real part of the mode, $\Omega=\omega_R$. The imaginary part is then related to the width of the resonance. In this picture, the moon serves as external force oscillating at the orbital frequency. A detailed computation shows that the height of the resonant flux (i.e. the absolute value of the horizon flux at the resonant frequency) \emph{grows} in the large distance limit~\cite{Cardoso:2011xi}
\begin{equation}
 |\dot E_{\rm resonance}|\equiv{\dot E}_H|_{\Omega=\mu_s}\sim \mu_s^{-1/3}\sim (v/c)^{-1}\,,\label{Epeak}
\end{equation}
Since the quadrupolar formula gives $\dot{E}_\infty\sim (v/c)^{10}$, in case of massive scalar resonances the resonant energy ratio $|{\dot E}_{\rm resonance}|/\dot{E}_\infty$ grows
at large distance and the energy dissipated through tidal polarization at the event horizon eventually dominate over the energy emitted in gravitational waves to infinity. 
This opens the intriguing possibility of existence of \emph{floating orbits}~\cite{Misner:1972kx,Press:1972zz}, i.e. orbits of \emph{zero radial acceleration}, where the (positive) energy flux emitted in gravitational waves is entirely compensated by a (negative) resonant polarization flux, so that the total energy flux is zero and the energy balance equation gives
%
%
%
\begin{equation}
 \dot E_p=-\dot E_{\rm total}=0\,,
\end{equation}
where $E_p$ is the particle binding energy. Note that the condition above is satisfied for frequencies slightly smaller than $\Omega=\mu_s$, because the negative flux at the horizon can indeed exceed the gravitational flux at infinity. However, because the width of the resonance is tiny in the small mass limit [cf. Eq.~(\ref{wRwI})], the floating orbit frequency and the resonant frequency are very close to each other, see Ref.~\cite{Cardoso:2011xi,Yunes:2011aa} for a detailed discussion. Therefore, at $\Omega\sim\mu_s$, the binding energy is constant and the moon effectively hovers in a floating orbit.
The existence of these peculiar orbits was extensively discussed in Refs.~\cite{Cardoso:2011xi,Yunes:2011aa} by solving Teukolsky's equation for a Kerr BH in a class of scalar-tensor theories. In fact, this phenomenon occurs in several extensions of Einstein's theory, including any scalar-tensor theory with a mass term in the scalar potential.

On the light of the connection between Teukolsky-based flux and tidal heating, we can interpret the results of Refs.~\cite{Cardoso:2011xi,Yunes:2011aa} in terms of ``tidal floating'': if the orbiting scalar charge interacts via a light massive field, it can resonantly raise tides at the horizon when its orbital frequency is close to the mass of the scalar interaction. In this case tidal heating is enormously amplified and its contribution dominates the net energy extraction.

As an example of the results of Refs.~\cite{Cardoso:2011xi,Yunes:2011aa}, in Figure~\ref{fig:superradiance2} we show the gravitational and scalar fluxes (normalized by the mass ratio $\nu=m_p/M$) obtained for massive Brans-Dicke theory with a very natural choices of the parameters: a central Kerr BH with spin $a=0.6 M$, a Brans-Dicke coupling $\omega_{\rm BD}\sim 10^6$ (which is two orders of magnitude larger than the current Cassini bound\footnote{This is a very conservative assumption because, if $\mu_s\gtrsim 10^{-16} {\rm eV}$, $\omega_{\rm BD}$ is unconstrained by current observations~\cite{Perivolaropoulos:2009ak,Alsing:2011er}.}) and $\mu_s M=0.01$, corresponding to $\mu_s\sim10^{-17} {\rm eV}$ for a massive BH with $M\sim 10^5 M_\odot$. We stress that the floating frequency occurs for a velocity which is slightly \emph{smaller} than that corresponding to the narrow peak shown by the scalar flux in the right panel of Figure~\ref{fig:superradiance2} [for details, cf. the inset of Fig.2 in Ref.~\cite{Cardoso:2011xi} and the corresponding discussion].

\section{Observing strong tidal heating in astrophysical black holes}
Can we observe BH tidal acceleration? And which kind of information would it encode?
A two-body system like that we discussed -- a supermassive BH and its small moon -- is one of the most promising targets for future space-based gravitational-wave detectors. As we discussed, BH tidal acceleration is impossible in Einstein's theory in four dimensions, thus the observation of strong tidal effects (like the tidal floating discussed above and in~\cite{Cardoso:2011xi,Yunes:2011aa}) is a smoking gun for physics beyond General Relativity.

At first order in the mass ratio, tidal floating is a \emph{monochromatic} source emitting gravitational waves roughly at constant frequency $f_{\rm GW}\sim 4\pi \mu_s$. More detailed computations~\cite{Cardoso:2011xi,Yunes:2011aa}, including the contribution from mass and spin angular momentum extraction from the BH, actually show that the floating timescale is not infinite, but can easily exceed the Hubble time. In practice, the moon would likely to spend a cosmological time orbiting close to a resonant frequency and this would result in emission of gravitational waves with a huge phase difference relative to General Relativity. Such a large dephasing would prevent detection of the source in a General Relativity template-based, matched-filtering gravitational wave search~\cite{Yunes:2011aa}, but it might be potentially detected through a continuous signal search.
On the other hand, a possible detection which is consistent with General Relativity can then be used to put constraints on the existence of ultra-light particles to unprecedented level~\cite{Yunes:2011aa}. For a supermassive BH of $M\sim10^5 M_\odot$ -- the most typical target for proposed space-based detectors~\cite{AmaroSeoane:2012km} -- these effects may be detectable (or discarded) for a scalar field of effective mass $\mu_s\lesssim 10^{17} {\rm eV}$, which is precisely the order of magnitude proposed in several scenarios~\cite{Arvanitaki:2009fg,Arvanitaki:2010sy}. 

BH tidal acceleration effects may also leave an imprint in optical observations. If a star floats at large orbital separation, its orbit could be optically resolved by current or near-future telescopes. Furthermore, the possibility of tidal floating suggests an alternative explanation of some bizarre phenomena related to the dynamics of stars orbiting supermassive BHs~\cite{arXiv:0808.3150,Madigan:2010pg}. In the last fifteen years, several experiments have independently observed a depletion of old stars close to the center of our galaxy. While several mechanisms have been advocated to explain this depression in the distribution of stars at about $0.1$ parsec, it is intriguing to notice that such barrier would roughly correspond to the tidal floating due to the tides raised by a scalar field of mass about $10^{-25}$ eV. Detailed studies in this direction are necessary and interesting.

Finally, a remarkable by-product of our analysis is that BH tidal acceleration can occur in higher dimensions. Indeed, equation~(\ref{tidesGeneral}) is still valid in $D=n+2$ dimensions and the gravitational-wave emission is suppressed by powers of $\Omega$~\cite{Cardoso:2002pa}. By generalizing our previous computation, we obtain the normalized quadrupolar contribution in $D$ dimensions:
\begin{equation}
\frac{|\dot{E}_{\rm H}|}{\dot{E}_{\infty}}\sim\left(\Omega^{D+1} r_0^{2D+2}\right)^{-1}\sim (v/c)^{-\frac{(D-5)(D+1)}{D-3}}\,,
\end{equation}
and, for any $D>5$, tidal acceleration is a dominant effect at large distance. 
Likewise, a generalization of the polarization acceleration formula, Eq.~(\ref{ratiocharge}), to $D$ dimensions reads
\begin{equation}
\frac{|\dot{E}_{\rm H}|}{\dot{E}_{\infty}}\sim (v/c)^{-\frac{(D-5)(D-1)}{D-3}}\,,
\end{equation}
which again shows, by very simple arguments, that tidal acceleration is the rule, rather than the exception, in higher dimensions. This surprising result was recently confirmed in a fully relativistic setting~\cite{Brito:2012gw}.

To conclude, we have established a simple connection between wave absorption from spinning BHs and tidal heating of the BH horizon. These two effects are strongly connected and they can be equivalently adopted to interpret and to predict some aspects of BH dynamics in General Relativity and in more general scenarios. The great advantage of this connection stands in the fact that tidal effects can be already understood at Newtonian level, whereas superradiant scattering from spinning BHs generically requires a more involved perturbative analysis at relativistic level. Furthermore, our results give further support to the BH membrane paradigm~\cite{membranep} and extend it to the class of phenomena we discussed in this paper. 

Tides teach us that apparently small effects can sometimes have unimaginable consequences. Just like it is hard to imagine that small bulges in its surface force the Moon to show us always the same face, so it is curious that tiny particles may tidally overcome the attraction of huge BHs and prevent other objects to be swallowed by them. Likewise, it is intriguing that, if such particles exist, their imprint would be more accessible by looking at the sky with telescopes and gravitational-wave detectors, rather than hunting for them at particle accelerators. 
If this proves to be correct, many adventures lie ahead for sailors of the spacetime sea.

\vspace{0.1cm}
\noindent
{\bf Acknowledgments.}
We thank Ana Sousa for preparing all the figures in this manuscript for us. We enjoyed fruitful discussions with Emanuele Berti,
Leonardo Gualtieri, Jan Steinhoff and Nicolas Yunes. We are particularly indebted to Kent Yagi for his detailed comments. We thank all participants of the
YITP-T-11-08 workshop on ``Recent advances in numerical
and analytical methods for black hole dynamics'', and the Yukawa Institute for Theoretical Physics at Kyoto University,
where parts of this work were completed.
This work was supported by the DyBHo--256667 ERC Starting Grant, the
  NRHEP--295189 FP7--PEOPLE--2011--IRSES Grant, and by FCT - Portugal through PTDC
  projects FIS/098025/2008, FIS/098032/2008, CTE-ST/098034/2008,
  CERN/FP/123593/2011. P.P. acknowledges financial support provided by
  the European Community through the Intra-European Marie Curie contract
  aStronGR-2011-298297.  

\section*{Appendix: on the existence of an ergoregion for a stationary and axisymmetric black hole}
In this appendix, we shall prove that a sufficient condition for the existence of an ergoregion in a stationary and axisymmetric spacetime is the existence of an event horizon,  provided the metric is regular, continuous and asymptotically flat. 
Let us consider the most general stationary and axisymmetric metric\footnote{We also require the spacetime to be invariant under the simultaneous transformation $t\to-t$ and $\varphi\to-\varphi$. This symmetry, also known as ``circularity condition'', implies $g_{t\theta}=g_{t\varphi}=g_{r\theta}=g_{r\varphi}=0$ (cf., e.g,. Chandrasekhar's book~\cite{Chandrasekhar:1985kt}). While the circularity condition follows from Einstein and Maxwell equations in electrovacuum, it might not hold true in modified gravities or for more exotic matter fields.}:
\begin{equation}
ds^2=g_{tt} dt^2+g_{rr}dr^2+2g_{t\varphi}dtd\varphi+g_{\varphi\varphi} d\varphi^2+g_{\theta\theta}d\theta^2\,, 
\end{equation}
where $g_{ij}$ are functions of $r$ and $\theta$ only. The event horizon is the locus $r_H=r_H(\theta)$ defined as the largest root of the lapse function:
\begin{equation}
N_{r=r_H}\equiv\left( g_{t\varphi}^2-g_{\varphi\varphi}g _{tt}\right)_{r=r_H}=0\,. \label{defhor}
\end{equation}
In a region outside (resp. inside) the horizon, $N$ is positive (resp. negative).

On the other hand, the boundary of the ergoregion, $r_{\rm ER}=r_{\rm ER}(\theta)$, is defined by
\begin{equation}
\left. g _{tt}\right|_{r=r_{\rm ER}}=0\,, \label{defergo}
\end{equation}
and $g_{tt}$ is negative (resp. positive) in a region outside (resp. inside) the ergoregion.
From Eq.~(\ref{defhor}) we get, at the horizon, 
\begin{equation}
 \left.g_{tt}\right|_{r=r_H}= \left.\frac{g_{t\varphi}^2}{g_{\varphi\varphi}}\right|_{r=r_H}\geq0\,,
\end{equation}
where, in the last inequality, we assumed no closed timelike curves outside the horizon, i.e. $g_{\varphi\varphi}>0$. The inequality is saturated only when the gyromagnetic term vanishes, $\left.g_{t\varphi}\right|_{r=r_H}=0$.
On the other hand, at asymptotic infinity $g_{tt}\to-1$. Therefore, by continuity, there must exist a region $r_{\rm ER}(\theta)$ such that $r_H\leq r_{\rm ER}<\infty$ and where
\begin{equation}
 \left.g_{tt}\right|_{r=r_{\rm ER}}=0\,.
\end{equation}
This proves that an ergoregion necessarily exists in the spacetime of a stationary and axisymmetric BH. As a by-product, we showed that the boundaries of the ergoregion (i.e. the ergosphere) must lay outside the horizon or coincide with it, $r_{\rm ER}\geq r_H$. In the case of a static and spherically symmetric spacetime, $g_{t\varphi}\equiv0$ and the ergosphere coincides with the horizon.

\vspace{-0.2cm}
\section*{References}

\end{document}